\documentclass{aa} 
\usepackage{graphicx}
\usepackage{natbib}
\begin{document}
\title{Multisite photometry of the pulsating Herbig Ae star V346 Ori\thanks{Based on data collected with the 
REM 0.6-m, Loiano 1.5-m, SARA 0.9-m and SNO 1.5-m telescopes}}

\author{
S. Bernabei\inst{1,2} \and
V. Ripepi\inst{3}\and
A. Ruoppo\inst{3} \and
M. Marconi\inst{3}\and
M.J.P.F.G. Monteiro\inst{4} \and
E. Rodriguez\inst{5}\and
T.D. Oswalt\inst{6} \and
S. Leccia\inst{3}\and
F. Palla\inst{7}\and
G. Catanzaro\inst{8} \and 
P.J. Amado\inst{5}\and
M.J. Lopez-Gonzalez\inst{5}\and
F.J. Aceituno\inst{5}
}

   \offprints{V. Ripepi}

   \institute{
INAF-Osservatorio Astronomico di Bologna, Via Ranzani 1,
40127 Bologna, Italy \\
\email{stefano.bernabei@oabo.inaf.it}
\and
Departimento de Astrof\'{\i}sica, Universidad de La Laguna, Avda.
Astrofisico F. S\'anchez sn, 30071 La Laguna, Spain \\
\and
INAF-Osservatorio Astronomico di Capodimonte,
Via Moiariello 16, 80131, Napoli, Italy \\
\email{ripepi@oacn.inaf.it, marconi@oacn.inaf.it, ruoppo@oacn.inaf.it, leccia@oacn.inaf.it}
\and
DMA-Faculdade de Ci\^encias and Centro de Astrof\'{\i}sica da
Universidade do Porto, Rua das Estrelas, 4150-762 Porto, Portugal
\email{Mario.Monteiro@astro.up.pt}
\and
Instituto de Astrof\'isica de Andaluc\'ia, CSIC, Apdo. 3004, 18080 Granada, Spain 
\email{eloy@iaa.es,pja@iaa.es,fja@iaa.es,mjlg@iaa.es}
\and
Florida Institute of Technology, Department of Physics and Space Science, 150 W Univ. Blvd., Melbourne, FL 32901-6988, USA
\email{toswalt@fit.edu}
\and
INAF-Osservatorio Astrofisico di Arcetri, Largo E. Fermi, 5, I-50125
Firenze, Italy \\
\email{palla@arcetri.astro.it}
\and
INAF-Osservatorio Astrofisico di Catania, Via S.Sofia, 78, I-95123, Catania, Italy \\
\email{gca@oact.inaf.it}
}

   \date{}

% \abstract{}{}{}{}{}
% 5 {} token are mandatory

  \abstract
   {The study of pulsation in Pre--Main--Sequence intermediate-mass
stars represents an important tool for deriving information on fundamental 
stellar parameters and internal structure, as well as for testing
current theoretical models. Interest in this class of variable 
stars has significantly increased during the last decade and about 30
members are presently known in the literature.}
   {We have constructed the frequency spectrum of the oscillations 
   in V346 Ori. We apply asteroseismic tools to these data 
to estimate the intrinsic parameters (mass, luminosity, 
effective temperature) of V346~Ori and to obtain information on
its internal structure.}
   {CCD time series photometry in the Johnson V filter has been
obtained for a total of 145.7 h of observations distributed over 36 
nights. The resulting light curves
have been subjected to a detailed frequency analysis using updated  
numerical techniques. Photometric and spectroscopic data have also been 
acquired to determine reliable estimates of the stellar properties. }
   {We have identified 13 oscillation frequencies, 6 of which  with higher
 significance. These have been compared with 
the predictions of non-radial adiabatic models. The resulting best fit 
model has a mass of 2.1$\pm$0.2 $M_{\odot}$, luminosity 
$\log{L/L_{\odot}}=1.37^{+0.11}_{-0.13}$, and effective temperature 
7300$\pm$200 K. These values are marginally consistent with 
the association of V346 Ori to Orion OB1a. Alternatively, 
V346~Ori could be placed at a slightly larger distance than 
previously estimated.}
{}

   \keywords{stars: variables:  $\delta$ Scuti  -- stars:  oscillations --
	   stars: pre-main sequence --   stars: fundamental parameters --
           stars: individual V346 Ori
               }

  \maketitle
%
%________________________________________________________________

\section{Introduction}

Pre--Main-Sequence (PMS) stars with masses larger than $\sim$1.5
M$_{\odot}$ that show variable emission lines are known as Herbig
Ae/Be stars \citep{herbig}.  Observationally, they are found within
star forming regions and show variable emission lines (especially
H$\alpha$) and strong infrared excess caused by the presence of
circumstellar material.  In addition, Herbig Ae/Be stars are
characterized by photometric and spectroscopic variability on time
scales of minutes to years, mainly due to photospheric activity and
interaction with the circumstellar environment
\citep[e.g.,][]{gahm,bohm1}.

Although recent theoretical work has improved our understanding of PMS
evolution \citep[see e.g.][]{palla93,dantona,swenson}, there remain
differences in the models owing to uncertainty in the treatment of
convection, of the input physics and the zero-point of the stellar
ages.  It is therefore desirable to find independent ways to constrain
PMS evolutionary tracks and, in turn, the internal structure of
intermediate-mass stars which are subject to dramatic changes during
the short-lived PMS contraction phase.

Asteroseismology of Herbig Ae/Be stars can in principle test PMS
models by probing their interiors.  It is now well established that
these stars cross the pulsation instability strip of more evolved
stars during contraction toward the main sequence, comprising a class
of variable stars called PMS $\delta$ Scuti stars
\citep[][]{kurtz,mp98,catala,ripepi05}.
%The existence of pulsating Herbig Ae stars was originally proposed by
%\cite{breger1} who discovered two candidates in the young open cluster
%NGC 2264. More than 20 years later, the suggestion was confirmed by
%\cite{kurtz} and \cite{donati} who observed $\delta$ Scuti-like
%pulsations in the Herbig Ae stars HR~5999 and HD~104237, respectively.\par

The first theoretical investigation of the PMS instability strip based
on nonlinear convective hydrodynamical models was carried out by
Marconi \& Palla (1998, hereinafter MP98) who calculated its topology
for the first three radial modes.  A subsequent theoretical work by
\citet{suran2} made a comparative study of the seismology of a
1.8 $M_{\odot}$ PMS and post-MS star. These authors found that the
unstable frequency range is roughly the same for PMS and post-MS
stars, but that some non-radial ($g$) modes are very sensitive to the
deep internal structure.  More recently, \citet{griga06}  have
produced a theoretical instability strip for PMS stars for the first
seven radial modes.
%In particular, it is possible to discriminate between the
%PMS and post-MS phase using differences in the oscillation frequency
%distribution in the low frequency range ($g$ modes). \par
Since the work by MP98, many new PMS $\delta$ Scuti candidates have
been observed and the current census of known or suspected candidates
includes about 30 stars \cite[see e.g.][]{ripepi05,zw08}.  However,
only a few of these have been studied in detail
\citep[e.g.][]{v351,bohm2,ipper}. Thus, the overall properties of this
class of variable stars are still poorly determined.

In order to perform reliable asteroseismological analysis, it is
critical to obtain data more accurate than that available in the
literature. In particular, it is necessary to measure as many excited
oscillations as possible with high precision on frequency and,
ideally, absence of aliases. Prior to space data which will become
progressively available from the MOST, COROT and possibly KEPLER
satellites, the only way we can satisfy these requirements is to carry
out multisite observations on promising asteroseismological targets,
i.e. stars which are multiperiodic. Starting in 2003, our group has
applied the multisite technique to two targets, V351 Ori \citep{v351}
and IP Per \citep{ipper}. The observational results obtained on these
stars have been recently interpreted using non-radial adiabatic
pulsation theory \citep[see][]{ruoppo}. As a continuation of our
ongoing project, here we concentrate on the Herbig Ae star V346~Ori.
This object is classified as a member of the Orion OB1a subassociation
\citep{h06}, or of its subgroup centered on 25 Ori \citep{b07}. It was
originally suspected to be variable on short time scale by \citet{marconi2000}.
  Subsequently, \citet{v346}, on the basis of
only two observing nights, confirmed the $\delta$ Scuti nature of the
light variation by measuring two highly significant frequencies of
pulsation at $f_1$=35.2$\pm$1.3 c/d and $f_2$=22.6$\pm$1.5 c/d and two
less significant frequencies at $f_3$=45.7$\pm$1.3 c/d and
$f_4$=18.7$\pm$1.3 c/d.  The measured amplitudes of the oscillations
in the $V$ band were $A_1=$2.9, $A_2=$2.8, $A_3=$1.2, and $A_4=$1.4
mmag. %suggesting the presence of non-radial pulsation.
To improve on these initial
results, we carried out: 1) an extensive photometric multisite
campaign to enlarge the number of observed mode of pulsation; and 2) a
spectroscopic analysis to determine the values of $\log g$, $T_\mathrm{eff}$
and $v\sin i$ of V346~Ori; 3) a comparison with nonlinear pulsation
models to constrain the position of V346~Ori in the HR diagram.

%The results of our study on V346 Ori will be presented in two papers.
%In the present one, we will describe the observations, data reduction
%and frequency analysis. In the next one, we will present the
%comparison with the prediction of non-radial models. \par

The structure of the paper is as follows: in Sect. 2 we describe the
observations and data reduction; in Sect. 3 we deal with the frequency
analysis; in Sect. 4 we report the results of the spectroscopic
analysis; in Sect. 5 we compare the measured frequencies with model
predictions and derive an estimate of the parameters of V346~Ori; and
in Sect. 6 we give some final remarks.

%-----------------------------------------------------------
   \begin{figure}
   \centering
   \includegraphics[width=9cm]{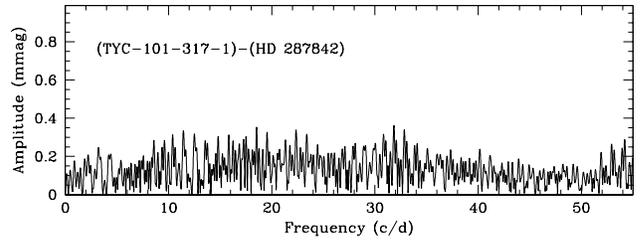}
      \caption{Fourier analysis of the differential photometry 
(TYC-101-317-1)-(HD 287842). See the text}
      \label{perconf}
   \end{figure}
%
%______________________________________________________________

%                                                One column figure
%-----------------------------------------------------------
   \begin{figure*}
   \centering
   \includegraphics[width=19cm]{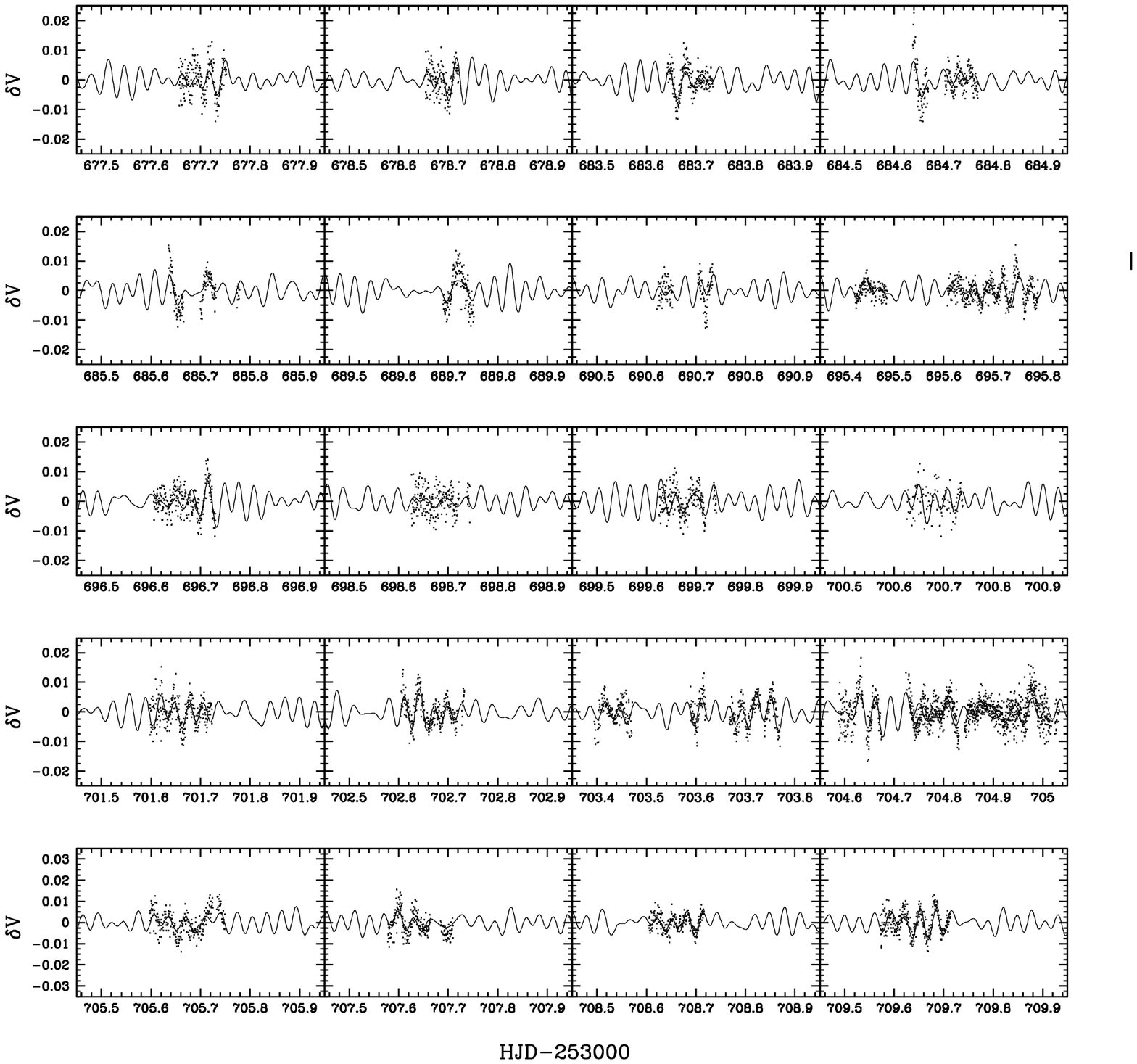}
      \caption{Light curves for  V346 Ori of the first 20 nights. 
Note that $\delta V\equiv V_{VAR}-V_{COMP}$, where VAR refers to V346~Ori and 
COMP is the comparison star HD~287842. 
In all panels, the solid line displays the fit to the
data with all the significant frequencies listed in Table~\ref{tab1} (see text).}
      \label{fig1}
   \end{figure*}
%
%______________________________________________________________

%                                                One column figure
%-----------------------------------------------------------
   \begin{figure*}
   \centering
   \includegraphics[width=19cm]{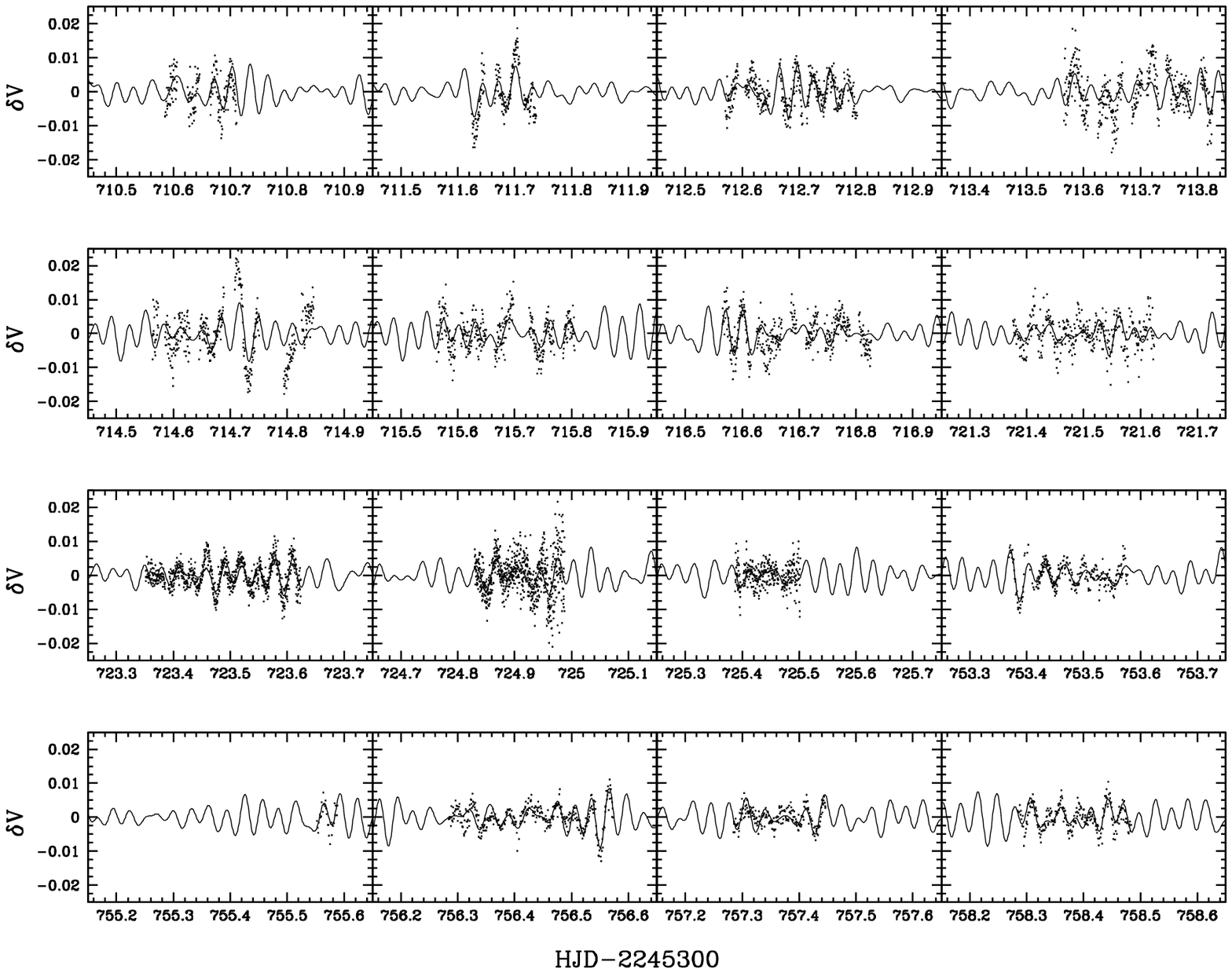}
      \caption{As in Fig. 1, but for the remaining 16 nights.}
      \label{fig2}
   \end{figure*}
%
%______________________________________________________________

\section{Observations and data reduction}

\subsection{Time series photometry}

Time series observations were obtained using four different telescopes
equipped with CCD devices during 36 nights between 2005, October and
2006, January. A complete description of the observations is given in
Table~\ref{tab1} (only available electronically).  As comparison star,
we observed HD~287842 (J2000: $05^h$ $24^m$ $24.21^s$ $+01^{\circ}$
$41^{'}$ $33.3^{''}$; V=10.3,  A7). This star was chosen because it is
only $\sim$5$^{'}$ away from V346 Ori and within the field of view of
all the instruments used.  Also, its magnitude and color are not too
different from those of V346~Ori.  The constancy of HD~287842 has been
checked with respect to other (fainter) stars present in the CCD
frames and in particular with TYC-101-317-1, a bright 
star (V$\sim$11.5 mag) which is present in the Sierra Nevada data 
(five nights, see Table~\ref{tab1} and below for the data analysis).
The Fourier analysis of the differential photometry\footnote{ Note that  
small nightly variations were removed by subtracting for each night the 
average photometry.} 
(TYC-101-317-1)-(HD 287842) is plotted in Fig.~\ref{perconf}. No significant 
frequency peak is present in the periodogram.
The observations were carried out 
only with the $V$ filter to have a time sampling sufficiently 
fast to sample properly the expected rapid pulsations. In the following we
describe the main features of the instrumentation used for this work
and give some details concerning the data reduction and photometric
procedures.

\begin{itemize}

\item
{\bf Loiano}\footnote{http://www.bo.astro.it/loiano/index.htm}: the
 observations were carried out using the 
1.5-m telescope equipped with the BFOSC instrument. The CCD
was a EEV 1340x1300 pixels with individual size of 0.58$^{''}$, for a
total field of view of 13$^{'} \times 13^{'}$. 
The data have been reduced following the usual procedures
(de-biasing, flat-fielding) and using standard IRAF routines. The
aperture photometry has been carried out using routines written in the
MIDAS environment. 

\item
{\bf REM}\footnote{http://www.rem.inaf.it}: the data were 
collected using the 0.60-m automated telescope at La Silla (Chile). 
Images were obtained with the ROSS spectrograph equipped with a 
commercial Apogee AP47 camera hosting a Marconi
47-10 1K$\times$1K 13 $\mu$m pitch CCD, which covers 
9.54$^{'} \times 9.54^{'}$ with a scale of 0.56$^{''}$/pixel. For the 
data reduction and photometry we followed the same recipes as for 
the Loiano observations. 

\item {\bf SARA}\footnote{http://saraobservatory.org} (Southeastern
Association for Research in Astronomy): we used the 0.9-m automated telescope at Kitt Peak
(USA).  The images were collected in the $V$ filter using an Apogee AP7p
camera with a back-illuminated SITe SIA 502AB 512x512 pixel CCD.
The pixels are 24 microns square, corresponding to 0.73$^{''}$ at the
telescope focal plane scale, giving a field of view of 6.2$^{'}
\times 6.2^{'}$. Sky flats, dark and bias exposures were taken every
night. The data were calibrated and reduced using standard IRAF
routines.

\item {\bf SNO}\footnote{http://www.osn.iaa.es} (Sierra Nevada
  Observatory): the data was obtained using the 1.5-m telescope equipped
  with a 2k$\times$2k CCD, with a pixel size of 0.23$^{''}$ for a
  total field of view of 7.92$^{'} \times 7.92^{'}$. Observations were
  carried out in the $V$ filter with exposures of about 7 s (depending
  on the quality of the night), using a 2x2 binning.  The data were
  reduced in a standard way by correcting the frames from bias and
  flat-field. Four different apertures were tested with radii of 4, 8,
  12 and 16 pixels. The 12 pixel aperture (corresponding to about
  5$^{''}$) was chosen as the optimal for our observations. The sky
  inner radius was fixed to 23 pixels and the annulus width to 3
  pixels.

\end{itemize}

In total, we collected 145.7 hr of observations during 36 nights,
distributed as follows: 18.7 hr with Loiano; 94.3 hr with REM ; 11.2
hr with SARA; 21.5 hr with SNO. \par

To prepare the data for Fourier analysis, we have first detrended 
the data to a common average zero value. This is needed 
for the following reasons: 
1) to avoid problems due to differences in the 
instruments+filters used; 2) to prevent problems with zero point 
differences between different datasets; 3) to eliminate the long term 
photometric variations due to the presence of circumstellar material 
around the star, a well known occurrence for Herbig Ae stars such as V346 Ori.
The detrending procedure consisted in removing a constant from the photometry 
for the majority of the nights. In the remaining cases, for nights with a duration of the  
observations larger than approximately 5 hours, we used a linear or quadratic regression
to detrend the data. As a consequence of this procedure, we are not able to investigate
frequencies lower than about 4.5-5 c/d.
The resulting light
curve of V346~Ori is shown in Fig.~\ref{fig1} and ~\ref{fig1}. Here, the solid lines
represent the least square fit to the data based on the frequency
analysis discussed in Section 3.

\subsection{Str\"omgren-Crawford photometry}

A few {\em  uvby$\beta$} points were also collected for V346~Ori in the Str\"omgren-Crawford 
photometric system for calibrating purposes using the six-channel 
{\em uvby$\beta$} spectrograph attached to the 0.9-m telescope at Sierra 
Nevada Observatory. During these measurements, three bright stars 
C1=HD~34888, C2=HD34745 and C3=HD~36525 were used as check stars, and 
instrumental magnitude differences were obtained relative to C1. To 
transform these instrumental differences into the standard 
{\em uvby$\beta$} system a set of 19 standards stars were selected from the 
list of \citet{Crawford1966} and \citet{Crawford1970}. 

Next, the absolute standard {\em uvby$\beta$} indices for V346~Ori and check 
stars were obtained following the method described in \citet{ridriguez2003}, using C1 to C3 as zero-points. The results are listed in Table~\ref{tab2new}
together with the values reported in the literature for the check stars. The error 
bars in this table mean standard deviations of magnitude differences relative 
to C1. As seen, our results are in very good agreement with the values 
found in the homogeneous catalogue of \citet{olsen96}. Similar results can be 
find in the lists of \citet{olsen93} and \citet{Hauck98}. 

These indices, together with suitable calibrations available in the literature 
for {\em uvby$\beta$}, can be used to estimate the physical parameters of 
V346~Ori using the procedure described in \citet{ridriguez2001}. In 
particular, a value of $\delta$m$_1$=0.012 is obtained for the metallicity 
parameter which leads to nearly solar abundances of [M/H]=$-$0.05($\pm$0.1) 
using the  \citet{smalley1993} relation for metal abundances. Moreover, using the 
grids of \citet{smalley1997} for solar abundances, values of 
$T_\mathrm{eff}$=7340($\pm$150)~K and $\log g$=4.0 ($\pm$0.1) are also found which are in 
very good agreement with those obtained from spectroscopy and discussed in Section 4.

\begin{table}
\caption{{\em uvby${\beta}$} indices obtained for V346~Ori and comparison 
stars. The pairs below the star names are the number of points collected for 
each object in {\em uvby} and $\beta$, respectively. The values given by 
\citet{olsen96} are listed in the bottom part.}
\label{tab2new}
\begin{tabular}{lrrrrr}
\hline
\noalign{\smallskip}
\multicolumn{1}{c}{Star} &   \multicolumn{1}{c}{\em V} &   \multicolumn{1}{c}{\em  b-y} & \multicolumn{1}{c}{\em m$_1$} &\multicolumn{1}{c}{\em    c$_1$} &  \multicolumn{1}{c}{$\beta$} \\
      &     (mag) &  (mag)       &   (mag)       &   (mag)      &   (mag)  \\
\noalign{\smallskip}
\hline
\noalign{\smallskip}
V346~Ori   &  10.173 & 0.169 &   0.180 &    0.811  &   2.773  \\
(5,5)      &       6 &     6 &       5 &       20  &      10  \\
C1=HD34888 &   6.773 & 0.165 &   0.167 &    0.923  &   2.758  \\
(6,6)      &       - &     - &       - &        -  &       -  \\
C2=HD34745 &   7.001 & 0.346 &   0.171 &    0.358  &   2.616  \\
(5,5)      &       5 &     3 &       4 &       12  &       4  \\
C3=HD36525 &   7.493 & 0.292 &   0.170 &    0.466  &   2.665  \\
(5,5)      &       7 &     8 &       4 &       12  &       4  \\
\noalign{\smallskip}
\hline
\noalign{\smallskip}
C1=HD34888  &  6.780 & 0.157 &   0.175 &    0.925 &      -   \\
C2=HD34745  &  7.001 & 0.351 &   0.164 &    0.362 &    2.614  \\
C3=HD36525  &  7.486 & 0.296 &   0.169 &    0.459 &    2.667  \\
\noalign{\smallskip}
\hline
\noalign{\smallskip}
\end{tabular}
\end{table}

\subsection{Spectroscopic data}

An echelle spectrum for V346~Ori was obtained with the BFOSC
instrument at the Loiano Telescope during the night of January 31,
2007.  We used the grism $\#9$ with the cross disperser $\#10$ in
order to improve the efficiency towards the blue. The resulting
spectrum includes 13 orders and covers the wavelength interval from
$\sim$3800 \AA~to $\sim$10000 \AA.  Data reduction has been
performed using the IRAF package ECHELLE. The resolution of our data
was R\,$\approx$\,2500.  The spectrum of
V346~Ori is shown and discussed in Section 4.

\section{Frequency analysis}

The pulsation frequency analysis has been carried out with the 
$period04$ package \citep{lenz} that adopts both Fourier and 
least-squares algorithms and permits simultaneously fitting of  
multiple sinusoidal variations, thus it does not rely on 
sequential prewhitening. In addition, $period04$ offers the possibility to 
weight the data during the least-squares procedure (see below).\par

\begin{table*}
\caption{Frequencies and amplitudes for V346~Ori calculated with $period04$ 
for the two cases, with and without weights (see text).
The estimated error on frequency is $\sim\pm$0.02 c/d (0.23 $\mu$Hz). For the no-weight case
the frequencies are listed in the same order of the weighted mode and not by decreasing amplitude.
 The horizontal line separates the high significant frequencies from the less significant ones.}
\label{tab2}
\begin{center}
\begin{tabular}{crccr|rccr}
\hline
\noalign{\smallskip}
\multicolumn{5}{c}{\bf weight case}  & \multicolumn{4}{c}{\bf no weight case}\\
\hline
%\noalign{\smallskip}
    &  Freq.  &Freq.    & Ampl. & S/N   & Freq. &Freq.    &   Ampl.  & S/N   \\
    &  (c/d)  &$\mu$Hz &(mmag) &       & (c/d) &$\mu$Hz &    (mmag)&     \\
%\noalign{\smallskip}
\hline
%\noalign{\smallskip}
$f_1$     &  35.107 & 406.33   & 2.23  & 12.1  & 35.107 & 406.33   & 2.20    & 8.8  \\
$f_2$     &  32.227 & 373.00   & 1.64  & 9.1   & 32.227 & 373.00   & 1.46    & 5.9  \\
$f_3$     &  31.611 & 365.86   & 1.18  & 6.6   & 31.611 & 365.86   & 1.11    & 4.6  \\
$f_4$     &  29.652 & 343.19   & 1.22  & 6.9   & 29.651 & 343.19   & 1.33    & 5.5  \\
$f_5$     &  29.159 & 337.49   & 0.86  & 4.9   & 29.159 & 337.49   & 1.08    & 4.5  \\
$f_6$     &  30.828 & 356.81   & 1.01  & 5.7   & 29.823 & 356.81   & 1.01    & 4.2  \\
%\noalign{\smallskip}
\hline
%\noalign{\smallskip}
$f_7$     &  24.833 & 287.42  & 0.82  & 4.6    &       &   &         &       \\
$f_8$     &  19.058 & 220.58  & 0.77  & 4.1    &       &   &         &       \\
$f_9$     &  12.936 & 149.72  & 0.77  & 4.1    &       &   &         &       \\
$f_{10}$  &  21.886 & 253.31  & 0.75  & 4.0    &       &   &         &        \\
$f_{11}$  &  27.446 & 317.66  & 0.78  & 4.5    &       &   &         &        \\
$f_{12}$  &  37.203 & 430.59  & 0.74  & 4.0    &       &   &         &        \\
$f_{13}$  &   8.045 &  93.11  & 0.69  & 4.0    &       &   &         &        \\
%\noalign{\smallskip}
\hline
\end{tabular}
\end{center}
\end{table*}

Before analysing the time series, we first note that the photometric
quality of the data varies among different nights and sometimes during
the same night. To account for this and to include the measured
photometric errors based on Poisson statistics which often are
underestimated with respect to the real scatter of the data, we have
used point-to-point weights for the frequency analysis.  Thus, we
adopted the weighting scheme in $period04$ called ``deviation
weight''. Defining $residuals$ as the absolute value of the difference
between observed and calculated phase points, this procedure assigns
the weights as follows: weight=1 for $residual <$ $cutoff$;
weight=[$(cutoff/residual)^2$] for $residual > cutoff$. The choice of
the $cutoff$ value is critical to obtaining reliable results. A
general rule is to avoid using a cutoff too small compared to the
amplitude of the oscillations \citep[e.g.][]{handler03}.  We found
that a $cutoff$ of 7.5 mmag complies with these requirements and
verified that lower values give unrealistic results, whereas higher
values (e.g $cutoff$=1 mmag) do not produce results significantly
different than the 7.5 mmag case. As described below, we performed the
frequency analysis in two ways: with and without weights.

%-----------------------------------------------------------
   \begin{figure}
   \centering
   \includegraphics[height=7cm]{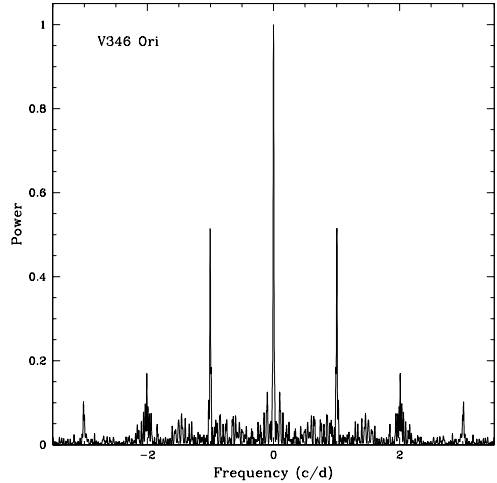}
      \caption{Spectral window (in power)}
      \label{fig3}
   \end{figure}
%
%______________________________________________________________

   The frequency analysis requires first the discussion of the
   Spectral Window (SW) which is shown in Fig.~\ref{fig3}.
   Unfortunately, due to the lack of an observing site in Asia, we
   were not able to completely cancel out the 1 c/d alias.
   Nevertheless, the SW of our data appears of good quality in both
   the weight and no-weight case, as the 1 c/d is reduced to about
   50\% (in power), and the 2 c/d alias is almost negligible.  After
   this, the procedure consists of the following steps: i) calculation
   of Fourier transform and identification of the highest amplitude
   peak; ii) least-square fit to the data with this frequency,
   improving simultaneously frequency, amplitude and phase; iii) check
   on the significance of the extracted frequency; iv) calculation of
   Fourier transform of the residuals and identification of the new
   highest amplitude peak; v) least-square fit to the data using
   simultaneously the derived frequencies; vi) iteration of the
   previous steps for the identification of all significant
   frequencies.  To determine whether to accept or reject a frequency,
   we adopted the widely used requirement of a signal to noise (S/N)
   amplitude ratio equal to 4 \citep{breger93}, corresponding to a
   value of about 12.6 in power \citep{breger2005}. In order to
   calculate the noise level, we used again $period04$ to obtain an
   average value for the noise itself in bins of 10 c/d width, moving
   along the spectrum. We verified that changing the size of the bins
   up to 50\% does not alter the S/N level significantly, especially
   for the low-amplitude frequencies close to the significance limit.

%We note also that the SW was taken into account for the step i), i.e.
%we verified that the alias pattern for each frequency was consistent
%with the one shown by the SW. In case of doubts concerning the correct
%frequency to fit, we choose the value which gave the best fit to the
%data. {\bf I DO NOT UNDERSTAND THIS SENTENCE}\par

\begin{figure*}
   \centering 
\includegraphics[width=19cm]{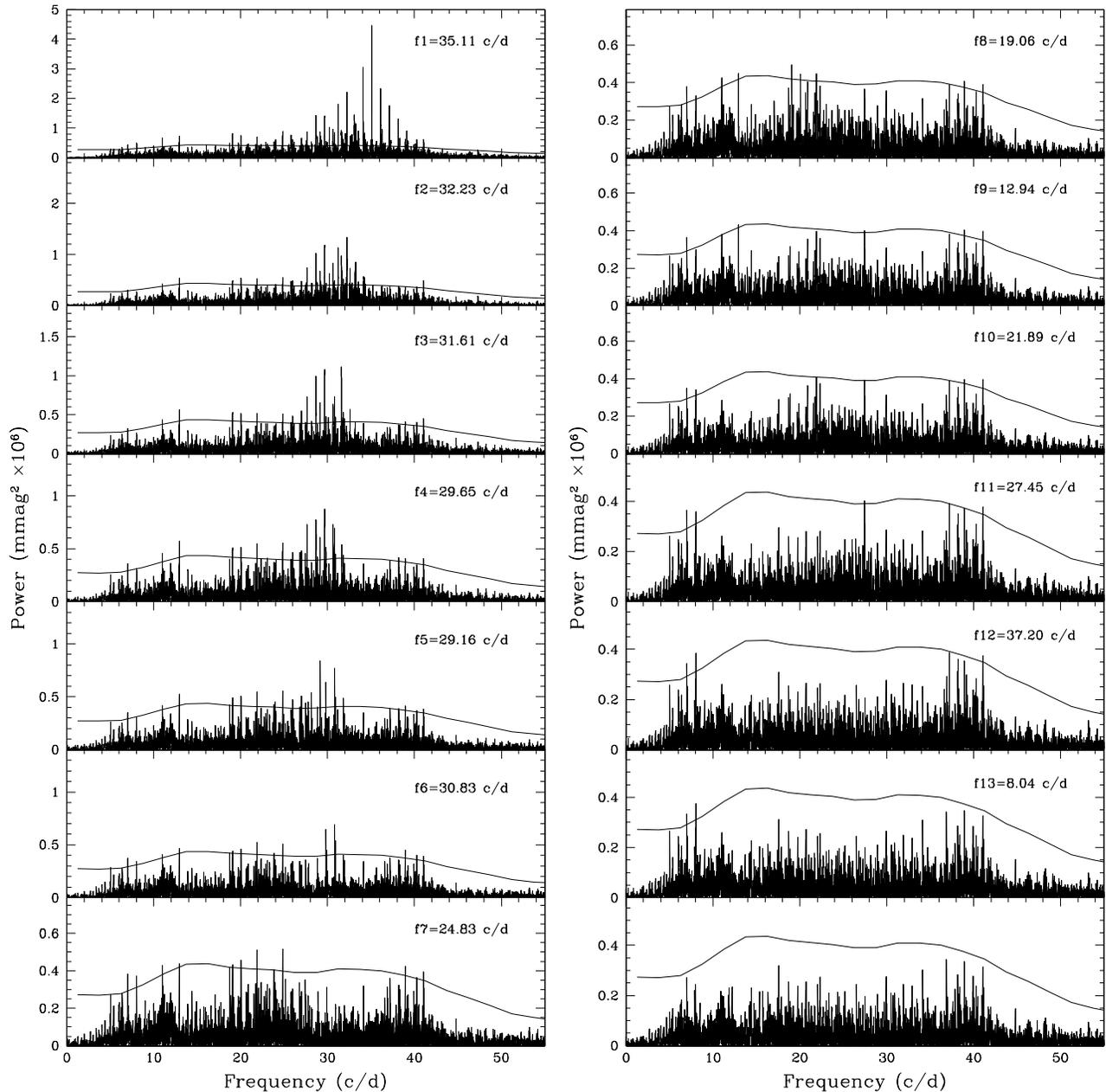}
      \caption{Frequency analysis for V346~Ori calculated 
with $period04$ adopting the weighting scheme of the data described in the text. 
The solid line shows the 99.9\% significant level, corresponding to 12.6 times 
the noise level. In each panel, one peak (i.e. the labelled
frequency) is selected and removed from the time series and a new
spectrum is obtained. Then, all the derived periodicities are fitted simultaneously to refine 
frequencies, amplitudes and phases. The analysis of the residuals produces the subsequent 
periodogram. The bottom figure in the right panel displays the periodogram after the
prewhitening with all the significant frequencies. }
      \label{fig4}
   \end{figure*}

The result of the frequency analysis is shown in Fig.~\ref{fig4},
while the derived frequencies are listed in Table~\ref{tab2} for
the weight and no-weight case. The difference in the two cases is
significant, since the number of derived frequencies varies from 13
in the weighted case to only 6 in the non-weighted case. However,
note that five of the six derived frequencies coincide in both
cases, while $f_6$ in the non-weighted model is clearly the -1 c/d
alias of $f_6$ in the weighted mode. As seen in Table~\ref{tab2},
the use of weights has the effect of reducing the amplitude peaks.
More significantly, the noise is also reduced, allowing more
frequencies to exceed the 4$\sigma$ threshold albeit only
marginally (see $f_7$ to $f_{13}$ in Table~\ref{tab2}). Given the poor 
significance of the frequencies derived with our weighting scheme, in the 
comparison with the models we shall use only the first six frequencies 
in Table~\ref{fig2}. The remaining ones are in any case useful to verify 
if the best-fit model is also able to reproduce them.    

Finally, we
have checked if some frequency could simply be a linear combination
of others and verified that all the detected periods are
independent of each other.

Regarding the errors on the extracted frequencies, 
we adopted the classical method of calculating the FWHM of the main
lobe in the SW, obtaining $\sigma f \sim$0.02~c/d.
%(see \citeal{alvarez}  for a discussion on the subject). 
 \citep[see ][ for a discussion on the subject]{alvarez}.  

%We are aware that the derived uncertainties 
%are overestimated, especially for the oscillation with largest S/N 
%frequencies. However, various formulations found in the literature 
%for the uncertainty on frequencies in time series data, 
%like \citet{mont99} and the recent one by \citet{kallinger08}, 
%provide underestimated errors as low as $\sigma f \sim$0.001~c/d 
%even in the worst case of S/N$\sim$4. 

%______________________________________________________________

%                                          One column figure
%-----------------------------------------------------------
%

%                                          Two column figure
%-----------------------------------------------------------
   \begin{figure*}
   \centering 
\includegraphics[width=18cm]{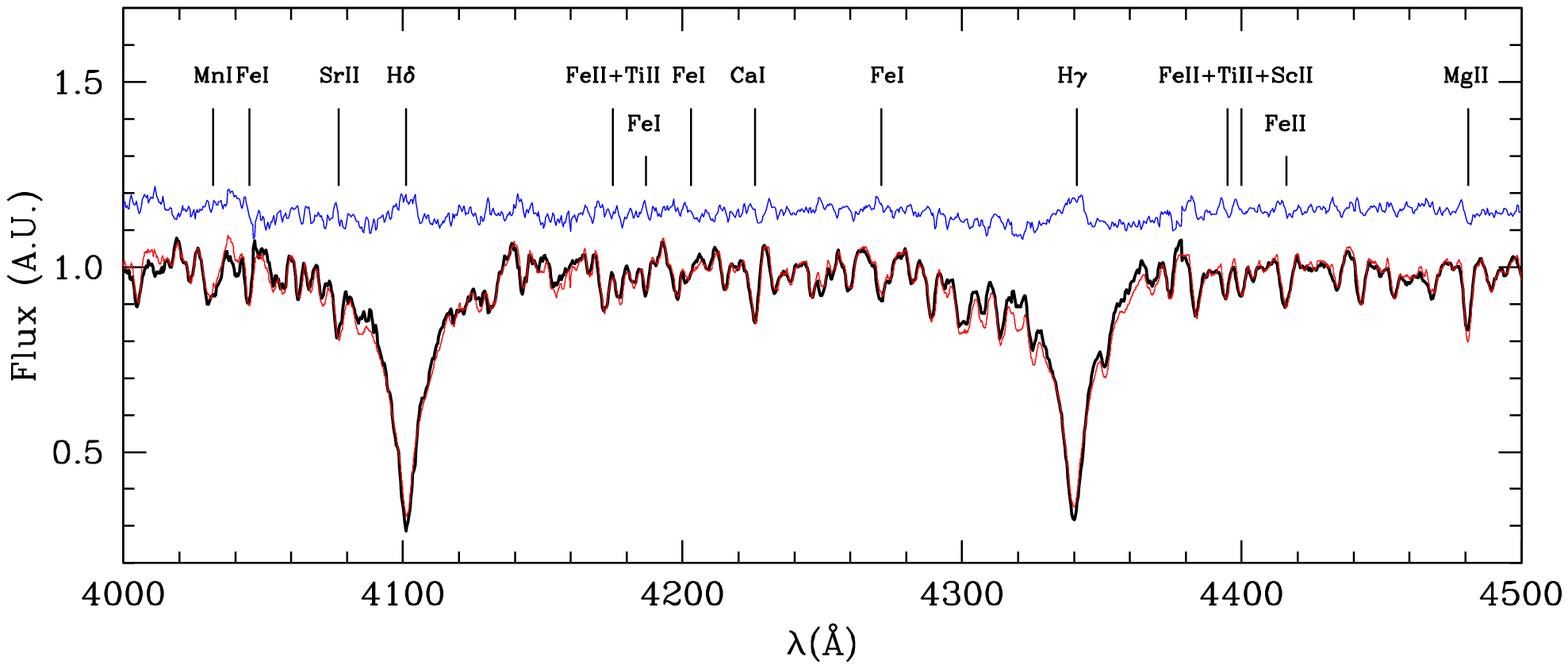}
      \caption{Selected regions of the normalized V346~Ori spectrum. 
The black and thin red and blue lines show the data for V346~Ori, for the comparison star 
HD124675 (A8IV) and the difference between the two spectra, respectively.}
      \label{fig5}
   \end{figure*}
%

%______________________________________________________________

%                                          Two column figure

    We can now compare the present results with those obtained by
   \citet{v346}. As a result, the two significant frequencies seen by Pinheiro et al. (2003) 
   are not in disagreement with our results.

\section{The spectrum of V346 Ori}

\subsection{Empirical determination of atmospheric parameters}

>From literature data we know that V346~Ori is
classified as a spectral type A5 or later and luminosity class III-V.
For our classification we followed the recipes described by \citet{gg89a, gg89b} 
for early F-type and late A-type stars,
complemented with additional criteria by \citet{jj}.

>From inspection of the spectral region
typically used for classification (e.g. \ion{Fe}{i} at $\lambda$4046 \AA;
\ion{Sr}{ii} at $\lambda$4077 \AA; \ion{Fe}{ii}-\ion{Ti}{ii} at $\lambda\lambda$4172-4179
\AA; \ion{Fe}{i} at $\lambda$4187 and 4203 \AA; \ion{Fe}{ii} at $\lambda$4415 \AA
/$\lambda$4481 \AA~ (\ion{Mg}{ii}) we conclude that V346~Ori is of spectral
type more similar to A9 than A7. Similarly, we tried to estimate the 
luminosity class of V346 Ori by analysing the 
morphology of the blends at $\lambda$$\lambda$4172-4179 \AA~and at
$\lambda$$\lambda$4395-4400 \AA~(\ion{Fe}{ii}-\ion{Ti}{ii}-\ion{Sc}{ii}) with respect to
$\lambda$4203 \AA~and $\lambda$4271 \AA~(\ion{Fe}{i}).  We computed the
ratio $\lambda\lambda$4172-4179/$\lambda$4271 and
$\lambda\lambda$4395-4400/$\lambda$4271, which is about 2. This means 
that V346 Ori luminosity class is IV or V \citep[see][]{gg89a}.  
 
\begin{figure}
\centering 
\includegraphics[height=8cm]{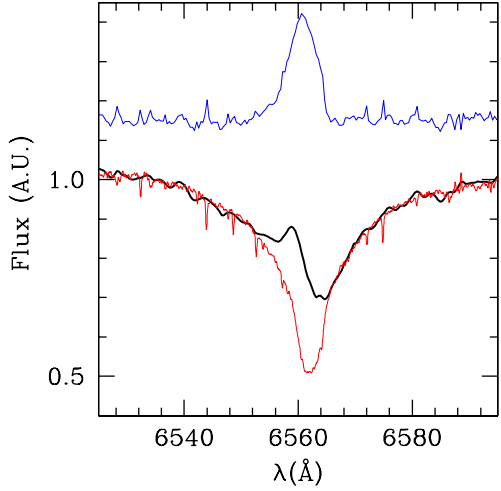}
\caption{As in Fig.~\ref{fig5}, but in the region of H$_{\alpha}$. Note 
the off-center emission present in the difference spectrum.}
\label{fig6}
\end{figure}

The following step was to determine more precisely the spectral type 
and simultaneously to estimate the rotational velocity (v$\sin$i) of V346 Ori. 
To this aim, we used the list of A-type stars with reliable and homogeneous 
v$\sin$i published by \citet{royer2002} coupled with high resolution spectra 
from the ELODIE archive \citep{moultaka04}. In particular we compared (visually) 
the spectrum of V346 Ori with several standard stars of spectral type 
A8 and A9 and luminosity class IV or V, as well as a variety of v$\sin$i. 
The best agreement was obtained with HD124675, an A8IV star with 
v$\sin$i$\sim$130 km/s and solar metallicity \citep[][]{fossati08,bush08}. 
Figure~\ref{fig5} shows the superposition of 
V346 Ori and HD124675 spectra in the region used for spectral 
classification, as well as the difference between the two spectra. 
Figure~\ref{fig6} show the region of H$\alpha$, which results partially 
filled and with an emission in the core due to the presence of hot gas 
around the star, a typical feature of Herbig Ae stars. \par 
We estimate that 
V346 Ori is an A8IV($\pm$ 1 spectral class) star with 
v$\sin$i$\sim$130 km/s and approximately solar metallicity.

This result agrees with some of the previous determinations, but
disagrees with others.  Considering spectroscopic measurements,
V346~Ori has been classified as A2IV \citep{mora}, A5IIIe
\citep{herbig}, A7III \citep{gc98}, A7III \citep{blondel06}, A8
\citep{valenti}, A8V \citep{vieira}, and A9 \citep{h06}\footnote{Note 
that some authors do not report the ``e'' after the spectral classification 
as it is usually done. We verified that this is not due to a 
lack of detection of emission (in H$\alpha$), but to various circumstances: 
e.g. the results by \citet{valenti} and \citet{blondel06} come from the UV etc.} 
Thus, our
estimate is consistent with the last three values, marginally with the
A7III evaluation by \citet{gc98} and \citet{blondel06}.
It is inconsistent with \citet{herbig} and \citet{mora}.  The
origin of this disagreement in spectral types cannot be easily
identified. We suggest two possible causes: 1) The modification of
the spectrum due to non-photospheric material placed along the line of
sight causing the star to appear earlier in some cases; and 2) The use of
different lines/methods for the classification. For example, using Ca
II K and/or H lines can often result in earlier spectral type
assignment for this type of stars \citep[see][]{gg89a,gc98}.
According to the temperature scale of \citet{kenyon}, our 
estimate for the spectral type of V346 Ori (A8$\pm$1 spectral class) 
corresponds to $T_\mathrm{eff}$=7580$\pm$250~K. Similarly, our estimated 
luminosity class IV-V correspond approximately to $\log g$=3.5-4.3
\citep{kaler}.

\subsection{Spectral synthesis determination of atmospheric parameters}

To validate our spectral classification and to determine the
physical parameters of V346~Ori, we have used synthetic spectra. 

The approach we used in this paper was to minimize the difference between observed and
synthetic H$_{\delta}$, H$_{\gamma}$ and H$_{\beta}$ profiles. As
goodness-of-fit test we used the parameter:

\begin{equation}
\chi^{2} = \frac{1}{N} \sum (\frac{I_{\rm obs} - I_{\rm th}}{\delta I_{\rm obs}})^{2}
\end{equation}

\bigskip

\noindent
where $N$ is the total number of points, $I_{\rm obs}$ and $I_{\rm th}$ are the intensities
of the observed and computed profiles, respectively, and $\delta I_{\rm obs}$ is the photon noise.
The synthetic spectra were generated in three  steps: first, we computed the stellar 
atmosphere model by using the ATLAS9 code \citep{kur93} then, the stellar spectrum 
was synthesized using SYNTHE \citep{kur81} and finally, the instrumental and 
rotational convolutions were applied. The ATLAS9 code includes the metal opacity by means of
distribution functions (ODF) that are tabulated for multiples of the solar metallicity
and for various microturbulence velocities. 
 
In this study we followed the method described in \citet{catanzaro08}, since 
the intersection of the three $\chi^2$ iso-surfaces is expected to improve the final
solution. To decrease the number of parameters, first we 
computed the $v_e \sin i$ of the star.
To derive the rotational velocity, we used SYNTHE, using a guess
model computed with the T$_{\rm eff}$ and $\log g$ estimated in
the previous sub-section, to reproduce the profile of
Mg{\sc ii}~$\lambda$4481 {\AA} line and obtained a value of 125~$\pm$~10~km~s$^{-1}$, 
in agreement with our previous estimate of $v_e \sin i \sim$~130~km~s$^{-1}$ .
For this star the best fit has been obtained for the model computed
with solar ODF for T$_{\rm eff}$\,=\,7550~$\pm$~200 K and $\log g$\,=\,3.5~$\pm$~0.4.

\begin{figure}
   \centering 
\includegraphics[width=9cm]{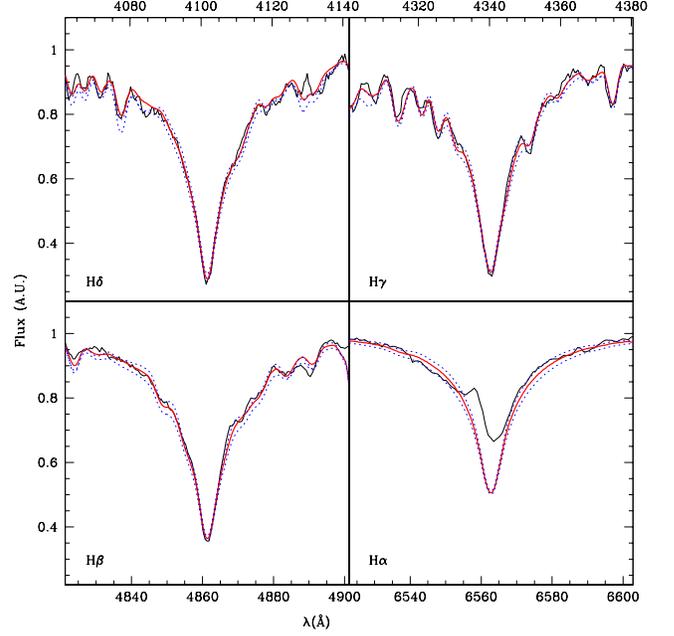}
      \caption{Zoom of V346 Ori's spectrum in the region of 
H$_{\delta}$, H$_{\gamma}$, H$_{\beta}$ and H$_{\alpha}$ in comparison with 
the best-fitting synthetic spectrum (T$_{\rm eff}$\,=\,7550 K, red line). 
Also shown in the figure for comparison are the synthetic spectra with 
T$_{\rm eff}\, \pm \, \delta\,T$ and $\log g\,\pm\, \delta\,\log g$ (blue dotted lines). 
Note that only 
H$_{\delta}$, H$_{\gamma}$ and H$_{\beta}$ were used for estimating the atmospheric 
parameter of V346 Ori; H$_{\alpha}$ is shown for comparison.}
      \label{fig7}
   \end{figure}

H$_{\alpha}$ is also present in our spectral range, we did not use it principally 
for the evident emission it shows in the core. The synthetic line overplotted
onto the observed one is shown in Fig.~\ref{fig7}. It has been computed by using
the atmospheric parameters derived from the other three Balmer lines. Also shown 
in the figure for comparison are the synthetic spectra with 
T$_{\rm eff}\, \pm \, \delta\,T$ and $\log g\,\pm\, \delta\,\log g$.\\

In conclusion, we have evaluated the atmospheric parameters of V346 Ori 
by using three different methods: 1) Str\"omgren-Crawford photometry; 
2) empirical determination of spectral and luminosity class; and  
3) comparison with synthetic spectra. These approaches 
give consistent results within the errors. 
Given that the comparison with synthetic spectra  is the most precise and 
physically based among the methods we used, in the following we will use 
the atmospheric parameters derived with this approach, 
but enlarging the error box in T$_{\rm eff}$ to $\pm$~250 K to take into account 
the results of methods 1) and 2).

%-----------------------------------------------------------
%

%-----------------------------------------------------------
%

%-----------------------------------------------------------
\begin{figure}
\centering
\includegraphics[height=9cm]{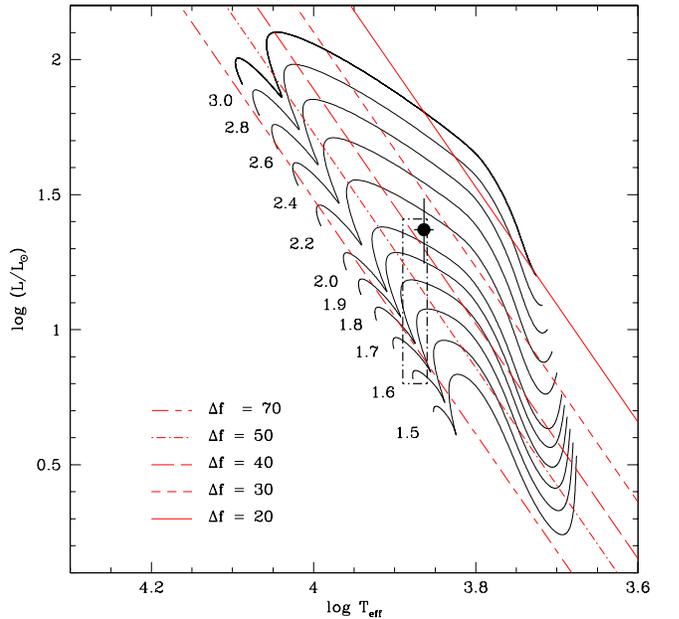}
\caption{HR diagram showing PMS evolutionary tracks computed
with the CESAM code for the labelled stellar masses (in
$M_{\odot}$). The dotted lines represent the large separation fit
obtained by \citet{ruoppo}. The rectangular box delimits the 
range of luminosity and effective temperature of V346~Ori from the 
literature. The best fit model is given by the filled dot.}  \label{fig8}
 \end{figure}
%  ______________________________________________________________

%-----------------------------------------------------------
   \begin{figure}
   \centering
   \includegraphics[height=9cm]{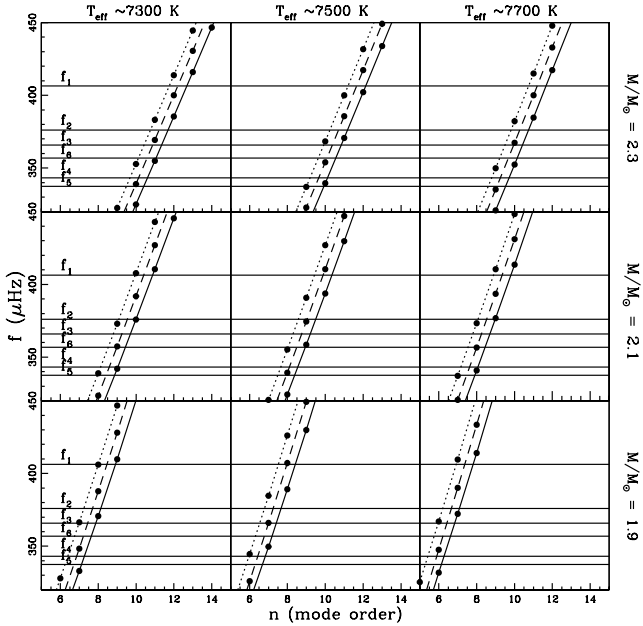}
   \caption{Predicted frequencies as a function of the radial order $n$. 
      Symbols are connected by a full line in the
      case of $l$=0, by a dashed line for $l$=1 and by a dotted 
      line for $l$=2. The horizontal lines represent the six most significant frequencies.}
  \label{fig9}
 \end{figure}
%  __________________________________________

\section{Comparison with models}

Having evaluated the physical parameters of V346~Ori, we now
 proceed to constrain its mass and  effective temperature (and, in turn,
its luminosity) comparing the observed frequencies with model predictions.
%\bf{For this purpose we first analyze only the six most 
%significant frequences.}  
In order to reduce the number of parameters in the theory versus observations 
comparison, we decided to restrict the analysis to the empirically 
estimated ranges of luminosity ($0.8 \la log L/L_{\odot} \la 1.41$) 
and effective temperature (7300-7800~K) for V346~Ori.
These ranges have been derived on the basis
of the atmospheric parameters estimated in the previous
section and are represented in  Fig.~\ref{fig8} by the dot-dashed box,
 together with the evolutionary tracks computed using the 
CESAM code \citep{morel}. From
the intersection between the box and the evolutionary tracks  we find that 
the predicted value of the stellar mass lies in the range from
 $\sim$1.6~M$_{\odot}$ to $\sim$2.2~$M_{\odot}$. For each stellar mass in 
this range, we selected
 three stellar models corresponding to three different effective 
temperatures, separated by $\sim$200~K, along the corresponding PMS 
evolutionary track, and performed the pulsation analysis in both radial and
 non-radial  modes (l=0,1,2) using the ADIPLS 
code\footnote{http://astro.phys.au.dk/~jcd/adipack.n/}. Additional models along 
the 2.3~M$_{\odot}$ have been computed in order to explore the behaviour of 
the predicted frequencies at higher stellar masses.
The predicted frequencies, for each selected stellar model, are used to 
reproduce the six most significant periodicities (from $f_1$ to $f_6$).   
The comparison between theory and observations is shown in Fig.~\ref{fig9}, 
where the frequencies are plotted as a function of the radial order for three 
assumptions l=0,1,2, varying the stellar mass and the effective temperature. 
Inspection of this plot, together with the search for the minimum  
$\chi ^2 $ \footnote{Given the observed frequencies of V346 Ori, we look 
for models, into the empirical box of  Fig.~\ref{fig8}, having 
$\chi ^2 \le 1$. As usual, the $\chi ^2 $ is defined as 
$ \chi^{2} = \frac{1}{N} \sum \frac{(f_{\rm obs} - f_{\rm th})^{2}}{\delta f_{\rm obs}^{2}+ \delta f_{\rm th}^{2}}$ 
where $N$ is the total number of observed frequencies, $f_{\rm obs}$ and $f_{\rm th}$ are the 
observed and the theoretical frequencies 
respectively, $\delta f_{\rm obs}$, $\delta f_{\rm th}$ are the corresponding error. 
If the program does not find any models with  
$\chi ^2 \le 1$, that means that not all frequencies are matched 
whithin the errors, it discards one by one the non-matched frequencies, 
computes again the $\chi ^2  $ for the remaining frequencies and searches 
again for models with  $\chi ^2 \le 1$. The model with the least value of 
the $\chi ^2 $  is the best-fitting model.} suggest that the 
best agreement is obtained for M$=$2.1~M$_{\odot}$, $\log{L/L_{\odot}}=$1.37
$T_\mathrm{eff}=$7314~K.  The position of the best fit model in the HR 
diagram is
represented by the large dot in Fig.~\ref{fig8}. The derived value of
the surface gravity of this model is $\log g$=3.8, close to our
empirical estimate of $\log g$=3.9$\pm$0.4. The various panels of Fig.~\ref{fig9} 
show that increasing or decreasing the stellar mass and/or the effective 
temperatures cause the discrepancies between observed and predicted frequencies 
to increase.
We have also verified that no reasonable match of the observed frequencies is 
found for stellar masses smaller than M$=$1.9~M$_{\odot}$.\\
 The comparison between the predicted and observed frequencies in the 
echelle diagram is presented in Appendix B, confirming that we find a combination
of mass, effective temperature and luminosity that is able to best reproduce 
the observed frequencies. We emphasize that this is true only within our 
explored range of stellar parameter and with the resolution in mass and 
effective temperature of the current model set

%In order to verify that the agreement improves as the stellar model mass increases 
%from $1.6 ~M$_{\odot}$ to $\sim$2.1~$M_{\odot}$,
%better agreement is obtained for still higher stellar masses we have 
%computed the frequencies also for models with M$=$2.3~M$_{\odot}$ and 
%shown that in this case we find a larger discrepancy with the observed 
%values (see the upper panels of Fig.~\ref{fig9}).}

 \section{Final remarks}

 We have presented the results of a multisite campaign on the
 pulsating Herbig Ae star V346~Ori. The frequency analysis has allowed 
us to extract six highly significant frequencies on the basis of a no-weight 
Fourier analysis, and additional seven less significant frequencies by adopting 
 a ``deviation weight'' scheme.
 The measured periodicities have been compared with the predictions of
 non-radial adiabatic models. We found that
 the resulting best-fit model has a stellar mass of 2.1$\pm$0.2~M$_\odot$,
 luminosity $\log{L/L_{\odot}}=1.37^{+0.11}_{-0.13}$, 
$\log g$=3.8$\pm$0.2, and effective temperature of $\sim$7300$\pm$200 K, 
where the errors are upper limits based
on the adopted mass and effective temperature steps. These estimates are in agreement 
with the spectroscopic estimates, within the associated uncertainties.

%{\bf Given the
% sensitivity of the echelle diagram topology to small variations in
% the input model parameters, the associated errors on the stellar
% properties of the best-fit model are found to be not larger than $\Delta
% M=$0.2~M$_{\odot}$, $\Delta \log$L=+0.0~L$_\odot$, and $\Delta
% T_\mathrm{eff}=$200~K. These upper limit errors  are 
%based on the adopted mass and effective temperature steps.}

The value of the asteroseismic luminosity is significantly higher than that
obtained by other authors. For example, Hern\'andez et al. (2005, H05 hereafter)
derived a value of $\log{L/L_{\odot}}=0.88$ (no errorbar is given)
using the average astrometric Hipparcos distance ($\sim$330$\pm$ 15 pc) to
the Ori OB1a association and the PMS evolutionary tracks by \citet{palla93},
 assuming $\rm A_V$=0 for V346 Ori. However, the uncertainties 
involved in this calculation are large. Indeed, to derive the luminosity of a star, even 
knowing the distance, one should know accurately the visual magnitude 
and the interstellar extinction. 
This is not straightforward for V346 Ori, which shows a wide variability in 
the visual band \citep[V=9.8-11.8;][]{the}, due to the presence of 
circumstellar material around the star \citep[see e.g. the SED published by][]{h06}. 
Similarly the value 
of $A_V$ to be used is not easily achievable: \citet{hernandez}  used 
$\rm A_V$=0, but the average absorption for the sub-group 25 Ori to which 
V346 Ori is believed to belong to \citep[see][]{b07} is $<A_V>=0.29$. 
It is useful to recalculate the $\log{L/L_{\odot}}$ value for V346 Ori adopting 
the distance by H05, but taking into account the value at minimum 
light for V346 Ori V=9.8 (which should be the closest to the photospheric one),
and a visual absorption $A_V$=0.41$\pm$0.1 as provided by the maps of
\citet{schlegel}, which are valid in an area of 6.1$\arcmin \times$ 
6.1$\arcmin$ 
around the star. As a result, allowing for an error of 5\% for the distance and 
10\% for both the visual magnitude and absorption, we get 
$\log{L/L_{\odot}}$=1.17$\pm0.11$, which agrees to within 1$\sigma$ with the results of 
our asteroseismological analysis. 
Of course, a better agreement could be obtained either by placing V346 Ori 
slightly beyond the Orion OB1a association (indeed the distance implied 
by the present results is D=400$\pm50$ pc that corresponds to the revised distance 
of the Orion Nebula Cluster, e.g. Jeffries, 2007) or supposing that the 
extinction due to circumstellar material has been underestimated. 
In the first case V346~Ori would not be a member of the Orion OB1a
 association. To our knowledge, the membership of V346~Ori to OB1a
has not been confirmed spectroscopically.  
To cast light on this complicated scenario, an
accurate measurement of the radial velocity of V346~Ori would be
highly desirable.

%________________________________________________________

\begin{acknowledgements}
V.R. wishes to thank the personnel of the Loiano Observatory for the
competent and kind help during the observations. S.B. acknowledges the
REM team for the large amount of observing time allocated for this
project and for the professional support during the
observations. Partial financial support for this work was provided by
PRIN-INAF 2005 under the project ``Stellar Clusters as probes of
stellar formation and evolution'' (P.I. F. Palla). This research has
made use of the SIMBAD database, operated at CDS, Strasbourg, France.
M.J.P.F.G.M. was supported in part by FCT through project {\scriptsize
  PTDC/CTE-AST/65971/2006} from POCI2010, with funds from the European
programme FEDER. T.D.O. was supported by NSF grant AST0206115.
P.J.A. acknowledges financial support from a Ramon y Cajal contract of
the Spanish Ministry of Education and Science.
\end{acknowledgements}

\bibliographystyle{aa}

\Online
\begin{appendix}
\section{Log of the observations}

\begin{table}
\caption{Log of the observations.}
\label{tab1}
\begin{tabular}{cccc}
\hline
\noalign{\smallskip}
HJD-2450000 & HJD-2453000  & Duration  & Telescope  \\
  start (days) &  end (days)&  (hours)  &      \\
\noalign{\smallskip}
\hline
\noalign{\smallskip}
 677.657 &677.752  &    2.3 &  REM \\
 678.654 &678.721  &    1.6 &  REM \\
 683.641 &683.734  &    2.2 &  REM \\
 684.638 &684.769  &    3.2 &  REM \\
 685.635 &685.779  &    3.4 &  REM \\
 689.690 &689.751  &    1.4 &  REM \\
 690.622 &690.734  &    2.7 &  REM \\
 695.421 &695.486  &    1.6 &  Loiano \\
 695.608 &695.791  &    4.4 &  REM \\
 696.605 &696.730  &    3.0 &  REM \\
 698.626 &698.744  &    2.8 &  REM \\
 699.626 &699.741  &    2.8 &  REM \\
 700.626 &700.738  &    2.7 &  REM \\
 701.598 &701.723  &    3.0 &  REM \\
 702.604 &702.733  &    3.1 &  REM \\
 703.395 &703.472  &    1.8 &  Loiano \\
 703.589 &703.769  &    4.3 &  REM \\
 704.587 &704.675  &    2.1 &  REM \\
 704.724 &705.032  &    7.4 &  SARA \\
 705.597 &705.750  &    3.7 &  REM \\
 707.578 &707.710  &    3.1 &  REM \\
 708.605 &708.717  &    2.7 &  REM \\
 709.572 &709.714  &    3.4 &  REM \\
 710.585 &710.711  &    3.0 &  REM \\
 711.625 &711.738  &    2.7 &  REM \\
 712.571 &712.802  &    5.5 &  REM \\
 713.566 &713.827  &    6.3 &  REM \\
 714.563 &714.847  &    6.8 &  REM \\
 715.563 &715.806  &    5.8 &  REM \\
 716.566 &716.827  &    6.3 &  REM \\
 721.376 &721.624  &    6.0 &  Loiano \\
 723.351 &723.624  &    6.6 &  Loiano \\
 725.388 &725.502  &    2.7 &  Loiano \\
 724.829 &724.987  &    3.8 &  SARA \\
 753.369 &753.577  &    5.0 &  OSN \\
 755.552 &755.588  &    0.9 &  OSN \\
 756.284 &756.573  &    6.9 &  OSN \\
 757.284 &757.446  &    3.9 &  OSN \\
 758.284 &758.483  &    4.8 &  OSN \\
\noalign{\smallskip}
\hline
\end{tabular}
\end{table}
\end{appendix}
%\Online
\begin{appendix}
\section{Theory versus observations in the echelle diagram}
\begin{figure*} \centering
   \includegraphics[height=10cm]{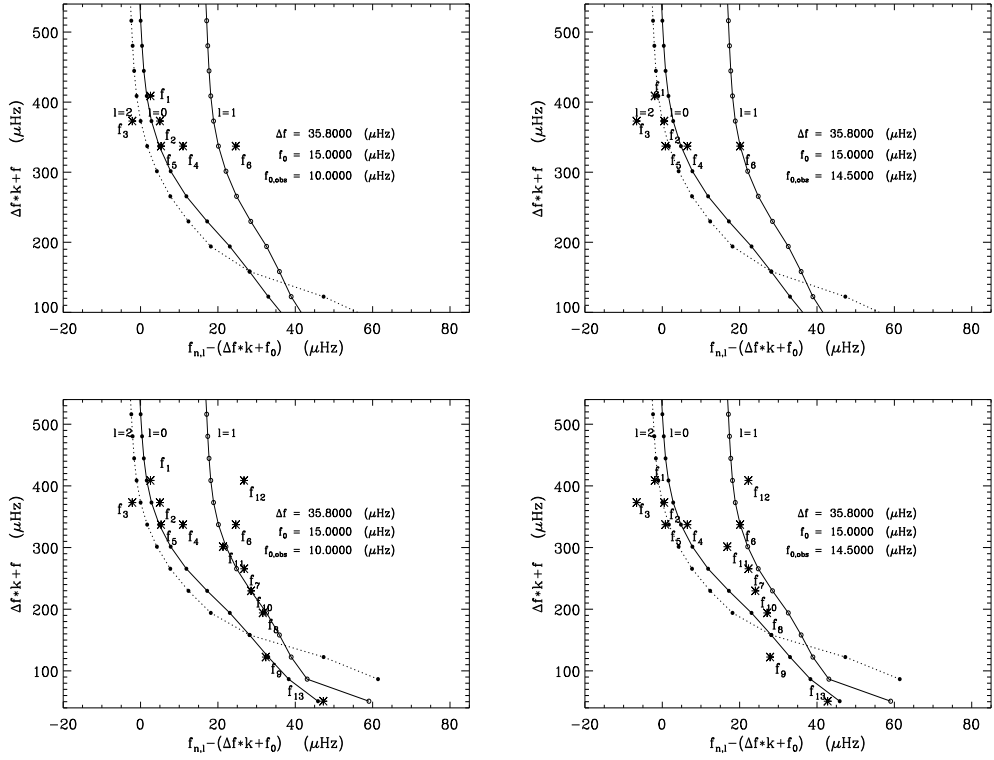}
   \caption{Echelle diagram for the model M=2.1 $M_{\odot}$,
     $T_\mathrm{eff}$=7314 K compared with the
     frequency data. 
The parameters adopted to match the data are
     labelled. Left: with the assumption
that the largest amplitude frequency ($f_1$) is radial (l=0) and
 by matching  only the six most significant frequencies (upper panel) 
or all the frequencies (bottom panel). Right: with the assumption that 
the largest amplitude frequency ($f_1$) is non radial (l=2) and by   
matching  only the six most significant frequencies (upper panel) 
or all the frequencies (bottom panel). }  \label{fig10} \end{figure*}

A powerful tool to investigate the prediction capability of 
pulsation models
is represented by the echelle diagram, where frequencies are reduced modulo 
$\Delta{f}$ by expressing them as
$f_{n,l}=f_0+\kappa \Delta f + \tilde{f} $
where $\Delta f$ is the large separation, $f_0$ is a suitably
chosen reference, and $\kappa$ is an integer such that $\tilde{f}$
is between 0 and $\Delta{f}$. The echelle diagram includes the information 
on the frequency large separation so that the comparison between modeled and 
observed periodicities in this diagram can allow us to constrain the 
properties of the frequency spectrum.
Hence, for all the models located in the empirical box of Fig.\ref{fig8}, 
we have used the computed frequencies to build a theoretical echelle diagram and 
search for the one that best reproduces the observed frequencies. 
As a result we obtain the best fit model quoted in Sect.9.
Looking at the behaviour of the best fit model in Fig.~\ref{fig9} 
we notice that the most significant frequency $f_1$ can be reproduced 
within $\pm$ 5$\mu$Hz either with l=0 or with l=2. Assuming l=0, as also 
supported by the idea that the largest amplitude frequency ($f_1$) is a 
radial mode, we obtain the echelle diagram shown in the left upper panel of 
Fig.~\ref{fig10} for a large frequency separation
$\Delta{f}=$35.8~$\mu$Hz. This quantity is evaluated using the
analytical relation between the large frequency separation, the
stellar luminosity and the effective temperature derived by \citet{ruoppo}.
As already found in Fig.~\ref{fig9}, the best fit model is able to reproduce the 
most significant frequencies within an uncertainty smaller than 5 $\mu$ Hz 
apart from $f_4$. This frequency is very close to $f_5$ and, as discussed 
by \citet{bregerbis}, close frequency pairs should be investigated accurately in
order to establish if they are real or not.

%Returning to the echelle diagram of Fig.~\ref{fig10}, note that the
%best fit model can reproduce all the observed periodicities with
%$l=0,1,2$ p-modes (with an uncertainty smaller than $5 \mu Hz$), with
%the exception of $f_4$, and $f_{13}$. We can try to invoke rotational
%splitting of frequencies to explain these last two periods that appear
%to deviate from the $l=1$ sequence by the same amount.  The frequency
%$f_4$ is very close to $f_5$ and, as discussed by Breger \& Bischof
%(2002), close frequency pairs should be investigated accurately in
%order to establish if they are real or not.  The frequency $f_{13}$
%could not be reproduced by our models, but in any case it is the least
%significant one.
 
As already specified, the best fit pulsation model has been found by 
requiring it to match only the six most significant frequencies. If the other 
seven measured frequencies (from $f_7$ to $f_{13}$) are added in the echelle 
diagram we obtain the plot shown in the left lower panel, where the only 
deviating frequency, beyond the already noted $f_4$, is $f_{12}$. 

From our spectroscopic analysis we have found that V346~Ori has a
projected rotational velocity of $v~sin i=125\pm$25~km/s.  The effect
of rotation on non-radial modes can be roughly estimated using the asymptotic
relation for non-radial displacements. This is given by $m f_{rot}$,
where $m$ is the spherical harmonic number and 
$f_{rot}\equiv \frac{vsin i}{2 \pi R sin i}$ is the rotational frequency.  For the
inclination angle $i$ we considered the two cases $i_1=\frac{\pi}{4}$
and $i_2 = \frac{\pi}{2}$, while for the radius we took the value of
our best fit model. Then, with these assumptions, the resulting rotational splitting is
$\Delta f_1 \simeq m 5.3$ $\mu Hz$ and $\Delta f_2 \simeq m 6.2$
$\mu$Hz, for $v sin i=$100 km/s and $\Delta f_1 \simeq m 7.9$~$\mu$Hz
and $\Delta f_2 \simeq m 9.3$~$\mu$Hz for $v sin i=$150 km/s.  
It is interesting to note that the displacement of $f_4$ and $f_{12}$, from
the $l=1$ sequence  seems to be consitent with the rotational splitting estimate 
for $v sin i=$150~km/s, with these frequencies corresponding to
$l=1$, $m= \pm 1$ modes. However, we are aware that this is a very 
simpified treatment of the rotational splitting phenomenon and 
that a more accurate analysis also  by means of 
asteroseismological models including rotation is needed in order to 
properly test this possibility.

If we discard the assumption that $f_1$ corresponds to a radial mode and 
explore the alternative possibility suggested by the behaviour of the best 
fit model in Fig.~\ref{fig9}, i.e. that $f_1$ is a l=2 mode, we obtain the plot 
shown in the right upper panel of Fig.~\ref{fig10}. In this case all six 
frequencies are reproduced within $\pm5$ $\mu$Hz, apart from $f_3$, that 
deviates a bit more from the l=2 sequence. However, if the other seven less 
significant frequencies are added,  we find that  theoretical predictions are 
not able to provide a good match for all the frequencies (see the right lower 
panel of the same figure).

%-----------------------------------------------------------
% ______________________________________________________________

\end{appendix}
\end{document}